\documentstyle[12pt,twoside,fleqn,espcrc1,psfig]{article}

\newcommand{\AmS}{{\protect\the\textfont2
  A\kern-.1667em\lower.5ex\hbox{M}\kern-.125emS}}

\title{
\vspace*{-4.5cm}
\begin{flushright}
DPNU-98-04\\
TMUP-HEL-9802\\
DUKE-TH-98-158\\
\vspace*{0.8cm}
\end{flushright}
Non-Central Heavy-Ion Collisions are the Place to Look for
DCC}

\author{M. Asakawa\address{Department of Physics, School of Science,
	Nagoya University, Nagoya 464-8602, Japan},
        H. Minakata\address{Department of Physics,
	Tokyo Metropolitan University, Minami-Osawa, Hachioji,
	                    Tokyo 192-0397, Japan}
	and
	B. M\"{u}ller\address{Department of Physics, Duke University,
		Durham, NC 27708-0305, U.S.A.}
}

\begin{document}
\maketitle

\pagestyle{empty}

\begin{abstract}
We give two reasons why we believe that 
non-central ultrarelativistic heavy ion collisions are the 
place to look for the disoriented chiral condensates (DCC). 
First, we argue that the most probable quench scenario for 
the formation of DCC requires non-central collisions. 
Second, we show by numerical simulations that strong electromagnetic 
fields of heavy ions can exert a surprisingly large effect on the 
DCC domain formation through the chiral anomaly. 
The effect again requires non-central collisions. Interestingly, 
the result of simulations is consistent with the formation of 
correlated two domains of the chiral condensate, which are aligned 
in space,
perpendicular to the scattering plane, 
but misaligned in isospin space.

\end{abstract}

\section{INTRODUCTION}

Recently, the possibility of the formation of chirally misaligned
domains in very high energy collisions of hadrons or heavy nuclei
has attracted a lot of interest. These domains are called
disoriented chiral condensates. A DCC domain is a coherent
excitation of the pion fields and corresponds to a local and coherent
rotation of the chiral order parameter of the QCD vacuum.
There have been considerable activities in both
experimental and theoretical research on DCC.

So far, analyses of high energy heavy ion collisions has focused on 
central collisions \cite{nayak97}. It is because
these collisions are thought to be the most effective ones in
establishing the conditions required for the chiral phase transition
in hadronic matter at high temperature or baryon density. 
Here we discuss the possible importance of non-central collisions of
ultrarelativistic heavy ions for the formation of DCC domains.
The reasons are twofold, as we discuss in the following.

\section{QUENCH}

Although the possibility of DCC has been pointed out on general
grounds \cite{anselm,bjorken1,blaizot},
the only mechanism known to date
for the formation of DCC domains is the quench mechanism 
proposed by Rajagopal and Wilczek \cite{rw93}.
Its essence is summarized by the important condition: 
At some initial moment, the matter must be out of equilibrium. 
A way of creating out-of-equilibrium field configurations in 
isospin space is to proceed through the following two-step process. 
1) First, create a thermally equilibrated and chirally symmetric 
phase of matter.
2) Then, let it cool down rapidly so that the configuration of the 
chiral fields remains around the symmetric point, while the effective
potential favors a state with spontaneously broken chiral symmetry.

Central collisions are where the initial conditions are expected 
to correspond to the highest energy density. However, at the same 
time they are where the characteristic time scales are largest,
because the heated volume is of maximal size.
For the purpose of creating field configurations which are out of
equilibrium in isospin space, a smaller time scale for the expansion 
of the system is better. In this sense,
central collisions are where the formation of DCC may be less likely.
On the other hand, it is less probable that chiral symmetry is
restored in very peripheral collisions at all.
Thus, we conclude that it is quite important to investigate
not fully central but also non-peripheral events in the DCC hunt.

\section{ANOMALY}

Ultrarelativistic heavy ions are sources of strong electric and
magnetic fields. When two high energy heavy nuclei collide and
the chirally restored phase is created, strong electric and magnetic
fields coexist. It was pointed out in Ref. \cite{mm96}
that the electric and magnetic fields affect the motion of
the chiral field through the Adler-Bell-Jackiw anomaly.
In the following argument, we use the linear sigma model whose
parameters are the same as those used in Ref. \cite{ahw95}.
The anomaly gives the following additional term to the effective
potential,
\begin{equation}
V_{\rm anomaly} = -\frac{\alpha}{\pi f_\pi}
\vec{E}\!\cdot\!\vec{H} \pi_3,
\end{equation}
where $\alpha$ and $f_\pi$ are the electromagnetic fine structure
constant and pion decay constant, respectively.
By using the point charge approximation, one can
express $\vec{E}\!\cdot\!\vec{H}$ as
\begin{equation}
\vec{E}\!\cdot\!\vec{H} = - \frac{2Z^2 e^2}{M}
\frac{\gamma^2}{R_1^3 R_2^3}(\vec{r}\!\cdot\!\vec{L}),
\end{equation}
where $Z$ and $M$ are the charge and mass of each colliding
nucleus (we are assuming that identical nuclei are colliding like at
RHIC), respectively, and $\vec{L}=\vec{b}\times M\vec{v}$ with
impact parameter $\vec{b}$, and
\begin{equation}
R_{1,2}= \sqrt{\gamma^2 (z\mp vt)^2 + 
\left ( \vec{r}_{\perp}\pm \frac{\vec{b}}{2} \right ) ^2 }.
\end{equation}
We have taken the collision point as the origin of the coordinate
system and the $z$ axis as the collision axis in the above 
expression. It indicates that the anomaly effect vanishes in 
very central collisions.
In non-central collisions, the anomaly effect vanishes on the
scattering plane, and $\vec{E}\!\cdot\!\vec{H}$ has definite signs 
in the upper and lower half-spaces, and they are opposite with 
each other.  

Since the anomaly effect is proportional to $\alpha^2$ and
the duration of heavy ion collisions is short at high energies,
the change of the average value of the $\pi_3$ field during
a collision is almost negligible and only $\dot{\pi}_3$, the 
conjugate field of $\pi_3$, is changed. Thus, the effect of
the anomaly is summarized as a quasi-instantaneous kick
to the $\dot{\pi}_3$ field which is coherent within the
upper and lower half-spaces and takes opposite signs in 
each half-space \cite{mm96}.

\section{SIMULATION}

To uncover the effect of the anomaly induced kick on
the formation of DCC, we have run dynamical simulations of
the linear sigma model.
The code is essentially the same as developed in Ref. \cite{ahw95}.
The details will be reported elsewhere \cite{amm98}.
We have implemented 
the effect of the kick in the initial 
condition on the conjugate field $\dot\pi_3$, providing a uniform 
shift that is opposite in sign in the upper and lower half-spaces
with regard to the scattering plane,
\begin{equation}\label{kick}
\langle{\mit\Delta}\dot{\pi}_3\rangle = {\rm sgn}(y) a_n m_\pi^2,
\end{equation}
where $a_n$ is a parameter and we have defined the $y$ axis
to be perpendicular to the scattering plane. In the following
calculation, we shall take 0.1 for $a_n$ in order to simulate the situation
expected to be realized at RHIC and take the quench initial condition
defined in Ref. \cite{ahw95}. In order to concentrate on the
effect of the anomaly kick to the time evolution of the system,
we assume in this paper that the size of the system is infinite in
the transverse directions, while we assume longitudinal boost 
invariance \cite{bjorken2}
in the longitudinal direction.
\vspace*{-0.5cm}
\begin{figure}[hbt]
\begin{minipage}[t]{45mm}
\psfig{file=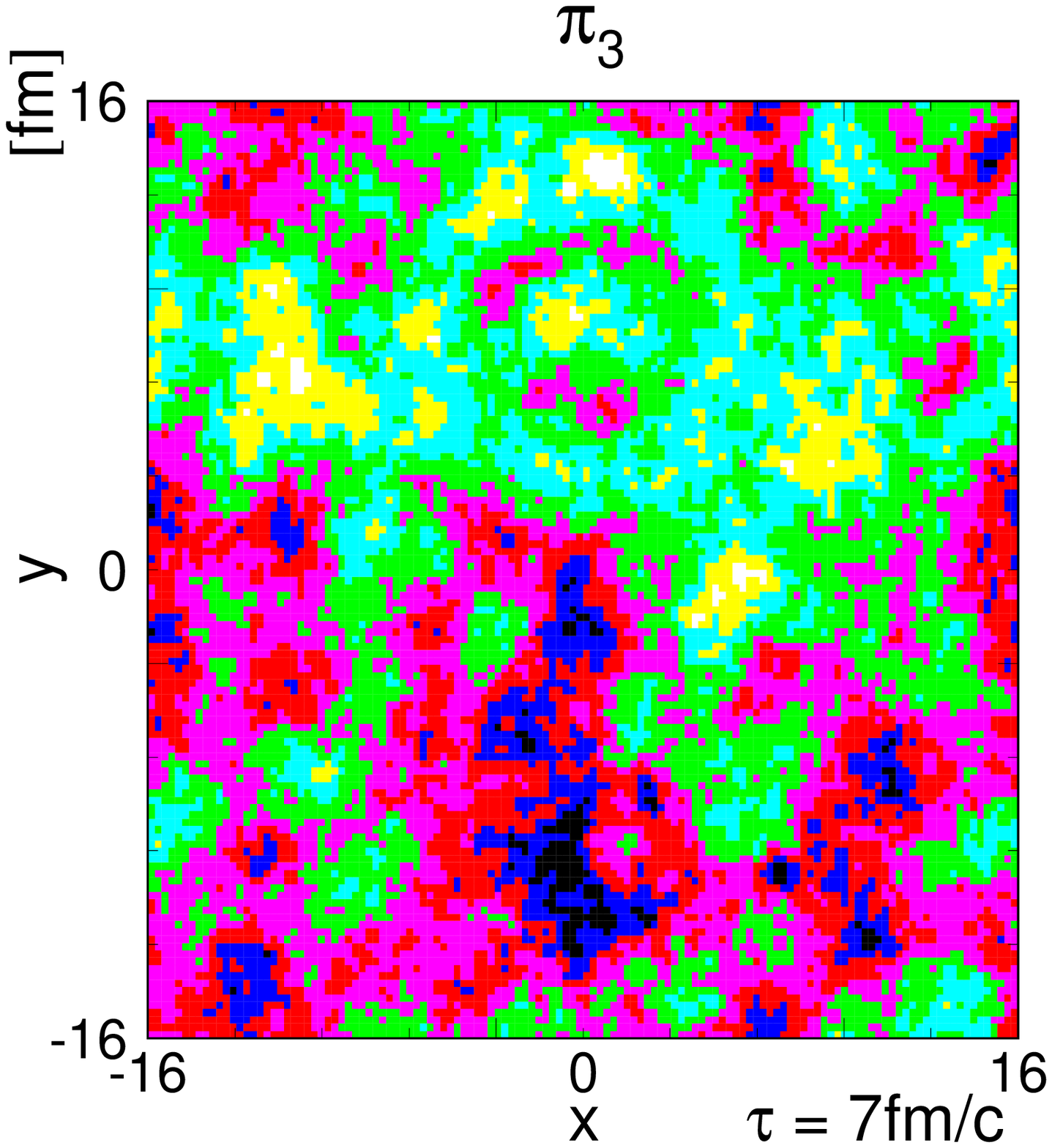,height=2.05in,angle=0}
\vspace*{-0.5cm}
\caption{Distribution of the $\pi_3$ field strength at $\tau = 7$ fm.}
\end{minipage}
\hspace*{0.1cm}
\begin{minipage}[t]{45mm}
\psfig{file=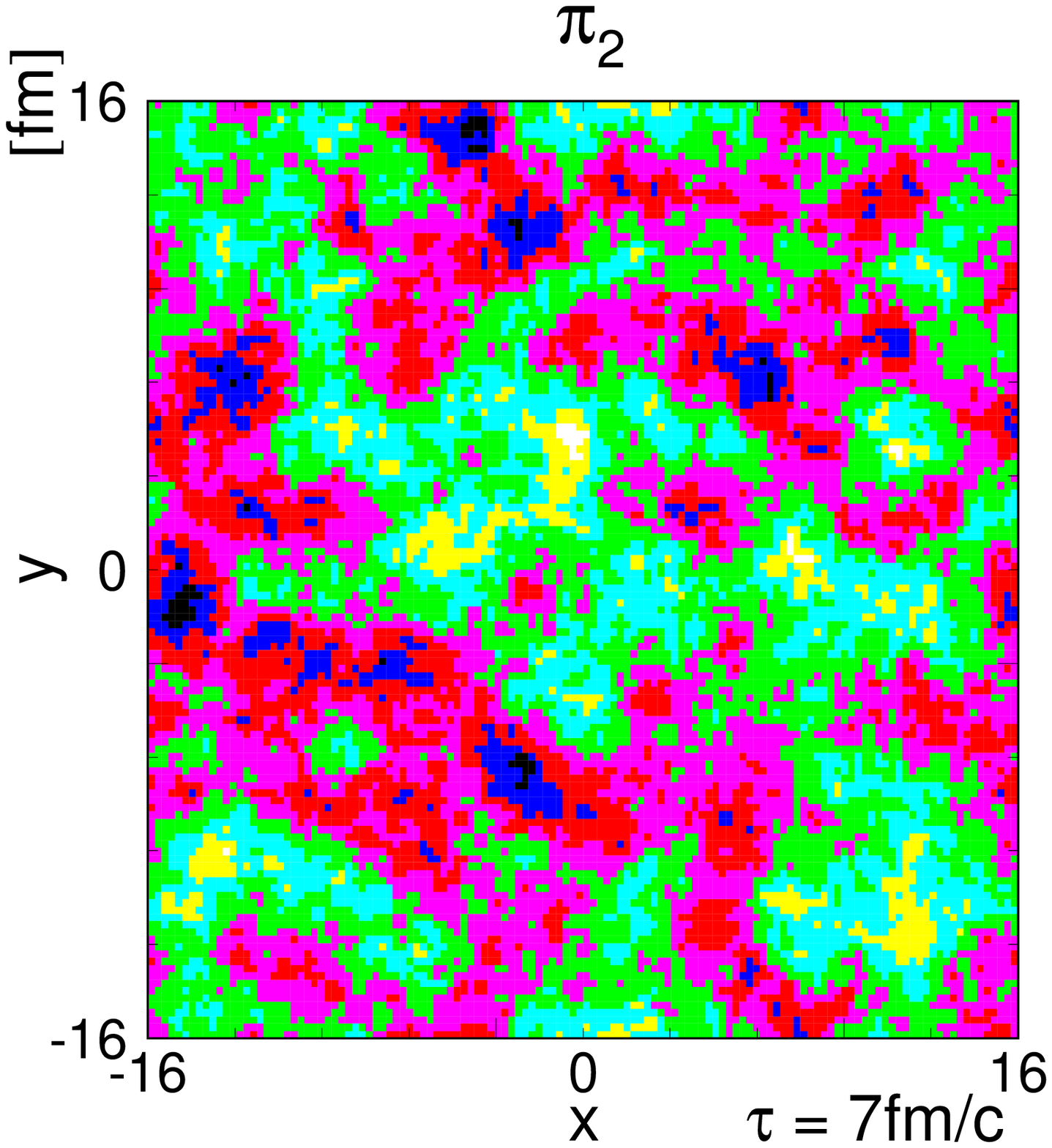,height=2.05in,angle=0}
\vspace*{-0.5cm}
\caption{Distribution of the $\pi_2$ field strength at $\tau = 7$ fm.}
\end{minipage}
\hspace*{0.1cm}
\begin{minipage}[t]{60mm}
\psfig{file=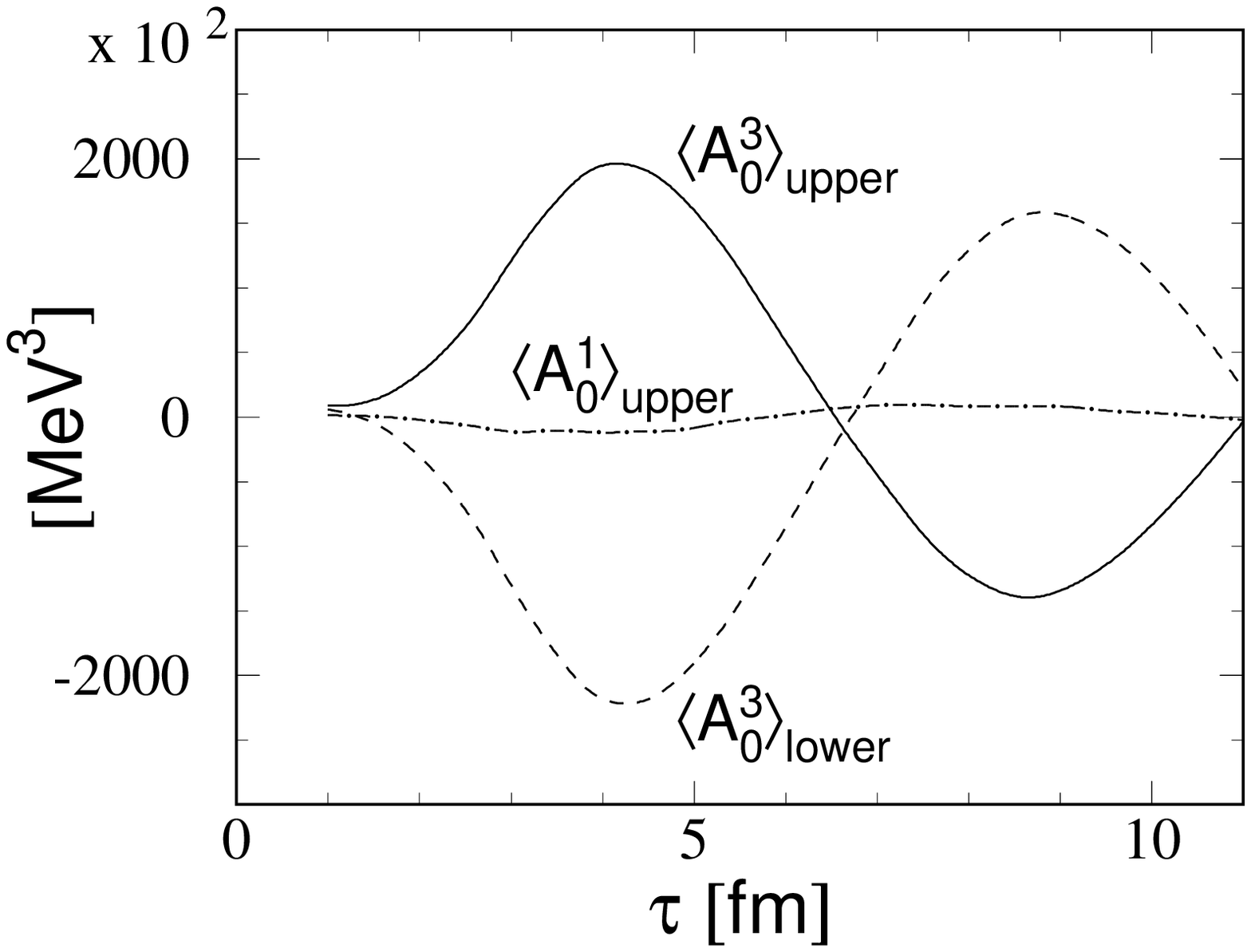,height=1.7in,angle=0}
\vspace*{-0.5cm}
\caption{Behavior of $A_0^1$ and $A_0^3$ averaged over the upper and
lower half-spaces and over 10 events.}
\end{minipage}
\end{figure}

\vspace*{-0.5cm}
Compared to the typical fluctuation of $\dot{\pi}_3$, the effect
of the anomaly kick given by Eq. (\ref{kick}) with $a_n = 0.1$
appears to be completely negligible. 
Our numerical simulations, however, show that a coherent kick
of such a small amplitude has a surprisingly large effect on
the time evolution of the $\pi_3$ field. In Figs. 1 and 2, we
show the distribution of the $\pi_3$ field, whose conjugate is 
kicked by the anomaly, and $\pi_2$ field, whose conjugate is
not kicked, respectively,
at $\tau=7$ fm in an event.
The initial proper time $\tau_0 $ has been taken to be 1 fm.
We can clearly see the asymmetry in the time evolution of the 
$\pi_3$ field between the upper and lower half-spaces, while 
no such asymmetry is observed in the $\pi_2$ field. This tells 
us that a kick of small magnitude induces a coherent motion of 
the $\pi_3$ field.

However, the pion field strengths themselves are not observables.
It is the currents, the vector $V_\mu^i$ and axial vector current
$A_\mu^i$, that couple to physical observables, where $\mu$ and
$i$ are Lorentz and isospin indices, respectively, and 
$V_\mu^i$ and $A_\mu^i$ are defined as
\begin{eqnarray}
V_{\mu}^i & = & \varepsilon^{ijk}\pi^j
\partial_{\mu}\pi^k, \nonumber \\
A_{\mu}^i & = &
\pi^i\partial_{\mu}\sigma
- \sigma\partial_{\mu}\pi^i.
\end{eqnarray}
We show in Fig. 3 the behavior of $A_0^1$ and $A_0^3$ averaged 
over the upper and lower half-spaces and over 10 events.
This vividly shows that striking coherence in $A_0^3$
throughout each of the upper and lower half-spaces so that
each can be considered as a DCC domain defined by the distribution
of $A_0^3$. On the other hand, $A_0^1$ does not show coherent
behavior. Neither does $V_0^1$ nor $V_0^3$. We have confirmed that
this is not affected by the finiteness of the system by
carrying out simulations with finite transverse dimensions.

\section{SUMMARY}

We have pointed out, on the basis of two reasons, the importance of 
semi-central high energy heavy ion collisions in the search for DCC. 
First of all, it is required by the quench scenario.
Secondly, it is preferred by the chiral anomaly in heavy ion
collisions.
$\vec{E}\!\cdot\!\vec{H}$ takes opposite signs in each side of 
the scattering plane and couples with the neutral pion field $\pi_3$
through the chiral $U(1)$ anomaly. Its effect can be summarized 
as a quasi-instantaneous kick to $\dot{\pi}_3$.
Its strength appears negligibly small, but our numerical simulations 
have indicated that the time evolution of the chiral order parameter 
is such that the anomaly-induced coherence in $\pi_3$ component is 
little affected by other field components in spite of their strong 
coupling at the Lagrangian level. 
$\pi_3$ and $A_0^3$ show definite asymmetry 
between the upper and lower half-spaces, and are highly coherent 
within each half-space. 

Thus, the anomaly induces the formation of DCC domains,
which are aligned in space, i.e., one in the upper half-space and
one in the lower half-space, but misaligned in isospin space.
Because of its definite sign, the spatial asymmetry could serve
as an additional experimental signal for the formation of DCC in
heavy ion collisions, in the context of an event-by-event analysis
with scattering plane identification, which is now experimentally
feasible.

\end{document}